\documentclass[10pt,a4paper]{article}

\usepackage{amsmath,amsthm,commath}
\usepackage{amssymb,textcomp}
\usepackage{cite}
\usepackage{tabularx}
\usepackage{graphicx}
\usepackage{caption}
\usepackage{subcaption}
\usepackage[T1]{fontenc}
\usepackage[utf8]{inputenc}
\usepackage{blindtext}
\usepackage{tabularx,ragged2e,booktabs,caption}
\usepackage{graphicx}
\usepackage{tikz}
\usepackage{enumerate}
\usepackage{algorithm}

\usetikzlibrary{shapes.geometric,arrows}
\newcolumntype{C}[1]{>{\Centering}m{#1}}

\newtheorem{theorem}{Theorem}
\newtheorem{lemma}[theorem]{Lemma}
\usepackage[english]{babel}
\usepackage[utf8]{inputenc}
\usepackage[noend]{algpseudocode}
\usepackage{epsfig}
\usepackage{color}
\usepackage{gensymb}
\usepackage{siunitx}
\usepackage{textcomp}
 \usepackage{graphicx}

\usepackage{graphicx}

\begin{document}
\title{Nonlinear State Estimation using Gaussian Integral}

\author{Kundan Kumar, and Shovan Bhaumik
\thanks{The authors are with the Department of Electrical Engineering, Indian Institute of Technology Patna, Patna 801103, India (e-mail: kundan.pee16@iitp.ac.in; shovan.bhaumik@iitp.ac.in)}
}

\maketitle

\begin{abstract}
		In this letter, a new filtering technique to solve a nonlinear state estimation problem has been developed. It is well known that for a nonlinear system, the prior and posterior probability density functions (pdf) are non-Gaussian in nature. However, in this work, they are assumed as Gaussian and subsequently mean, and covariance of them are calculated. In the proposed method, nonlinear functions of process dynamics and measurement are expressed in a polynomial form with the help of Taylor series expansion. In order to calculate the prior and the posterior mean and covariance, the functions are integrated over the Gaussian pdf with the help of Gaussian integral. The performance of the proposed method is tested in two nonlinear state estimation problems. The simulation results show that the proposed filter provides more accurate result than other existing deterministic sample point filters such as cubature Kalman filter, unscented Kalman filter, etc.
\end{abstract}



\section{Introduction}
Let us consider a system defined with the following state space model 	
\begin{equation}\label{state_eq}
x_{k+1}=\phi({x_k})+\eta_k,
\end{equation}
and 
\begin{equation}\label{measurement_eq}
y_k=\gamma({x_k})+\nu_k,
\end{equation}	
where $x_k \in \mathbb{R}^n$ is the state of the system, and $y_k \in \mathbb{R}^p$ is the output, $\phi(x_k)$ and $\gamma(x_k)$ are process and measurement functions.  $\eta_k \sim \mathcal{N}(0,\,Q_k)$ and $\nu_k \sim \mathcal{N}(0,\,R_k)$, are the process and the measurement noise respectively, which are assumed to be white, normally distributed and uncorrelated to each other. 

The objective of filtering is to determine the probability density function (pdf) of $x_k$ from the measurement data and approximate knowledge about the system. 
The most popular way of filtering is the Bayesian approach, where at each time $k$ the pdf of the state of $x_k$ is estimated recursively in two steps \cite{bhaumik2019nonlinear}: (i) prediction step (ii) update step. In the prediction step, the prior density function of state is calculated with the help of past measurements, $y_{1:k-1}$. This step is governed by the Chapman-Kolmogorov equation 
\begin{equation}\label{Chapman_kolmo_eqn}
p(x_k|y_{1:k-1})=\int p(x_k|x_{k-1}) p(x_{k-1}|y_{1:k-1}) dx_{k-1}.
\end{equation}
In the update step, Bayes' rule is used to compute the posterior density function of the state
\begin{equation}\label{Baye's_posterior}
p(x_{k}|y_{1:k}) \propto p(y_{k}|x_{k})p(x_{k}|y_{1:k-1}).
\end{equation}	

For a linear Gaussian system, the optimal solution of Eqs. \eqref{Chapman_kolmo_eqn}-\eqref{Baye's_posterior} exists, which is known as Kalman filter (KF) \cite{kalman1960new,bar2004estimation}. For nonlinear system, no optimal solution is available in general. Initially the extended Kalman filter (EKF) \cite{bar2004estimation,haug2012bayesian} was the solution, where a nonlinear system is linearized around the previous estimate, and then standard KF is used. However it diverges when linearization error is large.

Due to the limitation of the EKF, several other nonlinear filtering techniques have been developed. There are mainly two approaches; in one of them, a set of points in state space (also called particles) and their corresponding weights are used to represent the prior and the posterior pdfs. This approach is popularly known as the particle filter \cite{arulampalam2002tutorial}. Position and weights of the particles are updated in each iteration. Although usually, the accuracy of these filters is high, it has very high computational burden, and to some extent, it suffers from the `curse of dimensionality' problem.  

In another approach, the prior and the posterior pdfs (Eqs. \eqref{Chapman_kolmo_eqn}-\eqref{Baye's_posterior}) are approximated as Gaussian \cite{ito2000gaussian,wan2000unscented, arasaratnam2007discrete}, and represented with the mean and the covariance. The
prior mean $\hat{x}_{k+1|k}$ and the covariance $P_{k+1|k}$ of the prior estimate can be computed as follows:
\begin{equation}\label{prior_mean}
\hat{x}_{k+1|k} =\int \phi(x_k) \mathcal{N} (x_k;\,\hat{x}_{k|k},P_{k|k}) dx_k,
\end{equation}
\begin{equation}\label{prior_cov}
\begin{split}
P_{k+1|k}&=\int\phi(x_k)\phi^T(x_k)\mathcal{N}(x_k;\,\hat{x}_{k|k},P_{k|k})  dx_k \\& \hspace{0.8cm} -\hat{x}_{k+1|k}\hat{x}_{k+1|k}^T +Q_k.
\end{split}
\end{equation}
The mean and the covariance of the projected measurement can be calculated as 
\begin{equation}\label{mean_y}
\hat{y}_{k+1|k}=\int \gamma(x_{k+1}) \mathcal{N}(x_{k+1};\,\hat{x}_{k+1|k}, P_{k+1|k}) dx_{k+1},
\end{equation} 
\begin{equation}\label{cov_y}
\begin{split}
P^{yy}_{k+1|k}= &\int \gamma(x_{k+1}) \gamma^T(x_{k+1}) \mathcal{N}(x_{k+1};\,\hat{x}_{k+1|k}, P_{k+1|k}) dx_{k+1} \\&  -\hat{y}_{k+1|k} \hat{y}_{k+1|k}^T+R_{k+1}.
\end{split}
\end{equation}
The cross-covariance of the state and measurement is given as 
\begin{equation}\label{cov_xy}
\begin{split}
P^{xy}_{k+1|k}=&\int x_{k+1} \gamma^T(x_{k+1}) \mathcal{N}(x_{k+1};\,\hat{x}_{k+1|k}, P_{k+1|k}) dx_{k+1} \\& -\hat{x}_{k+1|k}\hat{y}_{k+1|k}^T.
\end{split}
\end{equation}
Finally, we compute the value of posterior mean and covariance as follows:
\begin{equation}\label{x_post}
\hat{x}_{k+1|k+1}=\hat{x}_{k+1|k}+K(y_{k+1}-\hat{y}_{k+1|k}),
\end{equation}
\begin{equation}\label{p_post}
P_{k+1|k+1}=P_{k+1|k}-KP^{yy}_{k+1|k}K^T,
\end{equation}
where $K$ is the Kalman gain,
\begin{equation}\label{Kalman}
K=P^{xy}_{k+1|k}(P^{yy}_{k+1|k})^{-1}.
\end{equation}
The integrals mentioned in Eqs. \eqref{prior_mean}-\eqref{cov_xy} are intractable for arbitrary $\phi(\cdot)$ and $\gamma(\cdot)$. Traditionally, they are approximated with the help of deterministic sample points, and associated weight. Some of the popular filtering techniques which use the above approach are the unscented Kalman filter (UKF) \cite{wan2000unscented,julier2000new, zheng2019unscented}, the Gauss-Hermite filter (GHF) \cite{arasaratnam2007discrete}, the cubature Kalman filter (CKF) \cite{arasaratnam2009cubature,wang2013spherical,bhaumik2013cubature,kumar2018higher}, \textit{etc.} As the above filters are suboptimal, there is a scope for improvement.   

In this letter, the authors proposed to use Gaussian integral \cite{straub2009brief} to evaluate the Eqs. \eqref{prior_mean}-\eqref{cov_xy}. Initially the function $\phi(x_k)$ and $\gamma(x_k)$ are assumed to be polynomial, if not they are expressed in a power series form \cite{apostol1969calculus,lang2012first} with the help of Taylor series. Each term of the power series is integrated over the Gaussian pdf with the help of Gaussian integral \cite{straub2009brief, stoof2009ultracold}. As a result, a new filtering technique has been evolved.  

The proposed method is applied to two nonlinear filtering problems. The accuracy is evaluated in terms of root mean square error (RMSE) and compared it with the EKF, the UKF and the CKF. It has been observed that the proposed filter is more accurate compared to the popular existing Gaussian filters.
 \section{Proposed Method of Filtering}
 \begin{lemma}
 	For any variable $y_1 \in \mathbb{R}$, the integral 
 	\begin{equation}
 	\begin{split}
 	I&=	\int_{-\infty}^{\infty} y_1^{m_1} \exp(-\frac{{y_1}^2}{2d_1}) dy_1\\ & = 0 \hspace{4cm} \text{if $m_1$ is odd,}\\&
 	=(2d_1)^{\frac{m_1+1}{2}} \Gamma(\frac{m_1+1}{2}) \hspace{1cm} \text{if $m_1$ is even,}
 	\end{split}
 	\end{equation}
 	where $d_1 \in \mathbb{R}^+$ and $m_1$ is any positive integer including zero. 
 	\begin{proof}
 		Let us consider an arbitrary function $f(y_1)=y_1^{m_1} \exp(-\frac{{y_1}^2}{2d_1})$. If $m_1$ is odd \textit{i.e.} $f(y_1)=-f(-y_1)$,
 		then the integral 
 		\begin{equation*}
 		I=\int_{-\infty}^{\infty} f(y_1) dy_1=0.
 		\end{equation*}
 		If $m_1$ is even \textit{i.e.} $f(y_1)=f(-y_1)$, then the desired integral
 		\begin{equation*}
 		I=2\int_{0}^{\infty} y_1^{m_1} \exp(-\frac{{y_1}^2}{2d_1}) \, dy_1.
 		\end{equation*}
 		By substituting  $y_1=\sqrt{2td_1}$, the above integral becomes
 		\begin{equation*}
 		\begin{split}
 		I&= (2d_1)^{\frac{m_1+1}{2}} \int_{0}^{\infty} t^{\frac{m_1-1}{2}} \exp(-t) \, dt\\&
 		=(2d_1)^{\frac{m_1+1}{2}} \Gamma(\frac{m_1+1}{2}).
 		\end{split}
 		\end{equation*}
 	\end{proof}
 \end{lemma}
 
 \textit{Corollary 1:} The above lemma can be easily extended for a variable $y \in \mathbb{R}^n$, where $y=[y_1,\, y_2,\, \dots, \, y_n]^T$, $d_i \, (i=1,2,\cdots,n) \in \mathbb{R}^+$ and $m_i$ is any positive integer including zero. Then the integral 	
 \begin{equation*}
 \begin{split}
 &\int_{\mathbb{R}^n} \prod_{i=1}^{n} y_i^{m_i} \exp\big( -\frac{1}{2} \sum_{i=1}^{n} \frac{y_i^2}{d_i} \big) dy  \\&
 \hspace{1.6cm}	=\prod_{i=1}^{n} (2d_i)^{\frac{m_i+1}{2}} \Gamma(\frac{m_i+1}{2}) \hspace{0.3cm} \text{when $m_i$ is even}, \\&
 \hspace{1.6cm}	= 0 \hspace{3.8cm} \text{otherwise}.
 \end{split}
 \end{equation*}
 
 \textit{Theorem 1:} For any arbitrary polynomial function $f(x)=\prod_{i=1}^{n}x_{i}^{m_i}$, the integral 
 \begin{equation}\label{Theorem_1}
 \begin{split}
 I&=\int_{-\infty}^{+\infty}f(x)\mathcal{N}(x;\mu,\,P)dx\\&=\frac{1}{\sqrt{(\pi)^n}}\bigg[\sum_{U_{a_{1j}}} \sum_{U_{a_{2j}}} \cdots
 \sum_{U_{a_{nj}}}  \prod_{i=1}^{n}  C_i \Big\{\mu_i^{a_{i1}}  \prod_{l=1}^{n}(S_{il})^{a_{il+1}}\Big\} \\& \hspace{0.5cm} \Big\{ (2d_i)^{(\frac{1}{2}\sum_{l=1}^{n}a_{li+1})}\Gamma\big(\frac{\sum_{l=1}^{n}a_{li+1}+1}{2}\big)\Big\}\bigg] \\& \hspace{1.7cm} \text{when $\sum_{l=1}^{n}a_{li+1}$ is even},\\&
 =0 \hspace{1cm} \text{otherwise}.
 \end{split}
 \end{equation} 
 Here, $a_{ij} \, (i=1,2,\cdots,n;\,j=1,2,\cdots,n+1)$ are positive integers including zero, $U_{a_{ij}}$ are all possible combinations of $a_{ij}$ which satisfiy $\sum_{j=1}^{n+1} a_{ij}=m_i$.  $d_i$ is the $i$-th eigenvalue of covariance matrix $P$, the orthogonal matrix $S$ ($|S|=1$) satisfies $S^{-1}PS = \text{diag}(d_1,d_2,\cdots,d_n)$ and $S_{il}$ are the element of $S$, and $C_i$ is the multinomial coefficient, which satisfies $ C_i=\frac{m_i!}{a_{i1}!a_{i2}!\cdots a_{in+1}!}$.
 \begin{proof} 	
 	Consider the integral 
 	\begin{equation}\label{integral_th1}
 	\begin{split}
 	I=&\frac{1}{\sqrt{(2\pi)^n|P|}} \int_{-\infty}^{\infty} \prod_{i=1}^{n}x_{i}^{m_i}   \exp \big\{-\frac{1}{2} (x-\mu)^T P^{-1} \\& (x-\mu) \big\}  dx,
 	\end{split}
 	\end{equation}	
 	where we substitute $x-\mu=Sy$. So $dx=|S|dy$, where $S$ is an orthogonal matrix which satisfies 
 	\begin{equation}\label{mat_sinvps}
 	S^{-1}PS = \text{diag}(d_1,d_2,\cdots,d_n),
 	\end{equation}
 	or
 	\begin{equation*}
 	S^{-1}P^{-1}S = \text{diag\big($\frac{1}{d_1},\,\frac{1}{d_2},\, \cdots, \frac{1}{d_n}$\big)}.
 	\end{equation*}
 	Now the integral \eqref{integral_th1} can be written as
 	\begin{equation*}
 	\begin{split}
 	I&=\frac{|S|}{\sqrt{(2\pi)^n|P|}} \int_{-\infty}^{\infty} \prod_{i=1}^{n} \big(\mu_i+\sum_{l=1}^{n}S_{il}y_l\big)^{m_i} \exp \big\{-\frac{1}{2} \\& \hspace{0.5cm} y^TS^TP^{-1}Sy  \big\} dy \\&
 	=\frac{1}{\sqrt{(2\pi)^n|P|}} \int_{-\infty}^{\infty} \prod_{i=1}^{n} \big(\mu_i+\sum_{l=1}^{n}S_{il}y_l\big)^{m_i}  \exp\big\{-\frac{1}{2} \\& \hspace{0.5cm}  \sum_{i=1}^{n}\frac{y_i^2}{d_i}\big\} dy .
 	\end{split}
 	\end{equation*}
 	With the help of multinomial expansion, the integral can be written as
 	\begin{equation*}
 	\begin{split}
 	I&=\frac{1}{\sqrt{(2\pi)^n|P|}}\int_{-\infty}^{\infty} \sum_{U_{a_{1j}}} C_1  \big(\mu_1^{a_{11}} \prod_{l=1}^{n} (S_{1l}y_l)^{a_{1l+1}}\big)  \sum_{U_{a_{2j}}} C_2 \\& \hspace{0.5cm} \big( \mu_2^{a_{21}} \prod_{l=1}^{n}(S_{2l}y_l)^{a_{2l+1}}\big)  \cdots  \sum_{U_{a_{nj}}}  C_n  \big(\mu_n^{a_{n1}}\prod_{l=1}^{n}(S_{nl}y_l)^{a_{nl+1}}\big) \\& \hspace{0.5cm}
 	\exp\big(-\frac{1}{2}\sum_{i=1}^{n}\frac{y_i^2}{d_i}\big) dy,
 	\end{split}
 	\end{equation*}
 	where $C_i=\frac{m_i!}{a_{i1}!a_{i2}!\cdots a_{in+1}!}$
 	is the multinomial coefficient, $U_{a_{ij}}$ consist of all possible  non-negative integer values of $a_{ij}$, which satisfies  $\sum_{j=1}^{n+1} a_{ij}=m_i$.  Now the above integral can be written as    	
 	\begin{equation*}
 	\begin{split}
 	I=&\frac{1}{\sqrt{(2\pi)^n|P|}}\int_{-\infty}^{\infty} \sum_{U_{a_{1j}}}\sum_{U_{a_{2j}}} \cdots
 	\sum_{U_{a_{nj}}} C_1 C_2 \cdots C_n  \\& \Big(\mu_1^{a_{11}}  \prod_{l=1}^{n}S_{1l}^{a_{1l+1}} \Big)  \Big(\mu_2^{a_{21}} \prod_{l=1}^{n} S_{2l}^{a_{2l+1}} \Big) \cdots 
 	\Big( \mu_n^{a_{n1}} 	 \prod_{l=1}^{n} S_{nl}^{a_{nl+1}} \Big) \\&
 	\Big(\prod_{i=1}^{n}y_i^{(\sum_{l=1}^{n}a_{li+1})} \Big)  	\exp\big(-\frac{1}{2}\sum_{i=1}^{n}\frac{y_i^2}{d_i}\big) dy.
 	\end{split}
 	\end{equation*}
 	Substituting $|P|=d_1d_2 \cdots d_n$ and using corollary 1 the above integral becomes 
 	\begin{equation*}
 	\begin{split}
 	I=&\frac{1}{\sqrt{(\pi)^n}}\bigg[\sum_{U_{a_{1j}}} \sum_{U_{a_{2j}}} \cdots
 	\sum_{U_{a_{nj}}}  \prod_{i=1}^{n}  C_i \Big\{\mu_i^{a_{i1}}  \prod_{l=1}^{n}(S_{il})^{a_{il+1}}\Big\} \\&  \Big\{ (2d_i)^{(\frac{1}{2}\sum_{l=1}^{n}a_{li+1})}\Gamma\big(\frac{\sum_{l=1}^{n}a_{li+1}+1}{2}\big)\Big\}\bigg] \hspace{0.1cm} 
 	\\& \text{when $\sum_{l=1}^{n}a_{li+1}$ is even},\\&
 	=0 \hspace{1cm} \text{otherwise}.
 	\end{split}
 	\end{equation*} 
 \end{proof}	 
 \textit{Illustration:} Let us consider a two dimensional system ($n=2$), and an arbitrary polynomial function $f(x)$$=$$x_1^2x_2 $. Here $m_1$$=$2 and $\sum_{j=1}^{3}a_{1j}$$=$$a_{11}$+$a_{12}$+$a_{13}$=2. 
 The possible values of  $\{a_{11},a_{12},a_{13}\}$=\{2,0,0\};\{0,2,0\};\{0,0,2\};\{1,1,0\};\{1,0,1\};
 \{0,1,1\}; and $m_2=1$, then $ \sum_{j=1}^{3} a_{2j}=a_{21}+a_{22}+a_{23}=1$. 
 Therefore the values of $\{a_{21},a_{22},a_{23}\}$ are \{1,0,0\};\{0,1,0\}; \{0,0,1\}.
 With the help of Theorem 1, the integral
 \begin{equation*} 
 \begin{split}
 I&=\int_{\mathbb{R}^2} f(x) \mathcal{N}(x;\mu,P)dx\\&
 =\big[\mu_1^2 \mu_2 + \mu_2 S_{11}^2 d_1+\mu_2S_{12}^2 d_2+2\mu_1S_{11}S_{21}d_1\\&\hspace{0.3cm}+2\mu_1S_{12}S_{22}d_2\big].
 \end{split}
 \end{equation*}   
	\subsection{General algorithm for proposed filtering}
	The algorithm of the proposed filter in brief is as follows:	 
	\begin{itemize}
		\item Initialize the filter with $\hat{x}_{0|0}$ and $P_{0|0}$.
	\end{itemize}
	\texttt{Step 1: Time update}
	\begin{itemize}
		\item Calculate the eigenvector matrix ($S$) and the eigenvalue matrix $(d)$ of $P_{k|k}$.
		\item Compute prior mean $\hat{x}_{k+1|k}$, and covariance $P_{k+1|k}$ by using Eqs. \eqref{prior_mean}-\eqref{prior_cov}, which are evaluated with Theorem 1.
	\end{itemize}
	\texttt{Step 2: Measurement update}	
	\begin{itemize}
		\item Again eigenvector matrix $(S)$ and eigenvalue matrix ($d$) of $P_{k+1|k}$ are calculated.
		\item Calculate $\hat{y}_{k+1|k}$, $P_{k+1|k}^{yy}$ and $P_{k+1|k}^{xy}$ by evaluating Eqs. \eqref{mean_y}-\eqref{cov_xy} with the help of Theorem 1.
		\item Compute posterior state estimate $\hat{x}_{k|k}$, and covariance $P_{k+1|k+1}$ by using Eqs. \eqref{x_post}-\eqref{Kalman}. 
	\end{itemize}	
	\section{Simulation Results}
	Problem 1. Here a single dimensional problem \cite{ito2000gaussian} has been considered, whose state dynamics and measurement equation are in the form of Eqs. \eqref{state_eq}-\eqref{measurement_eq}, with
	 $\phi(x_k)=x_k+5\delta t x_k(1-x_k^2)$ and $\gamma(x_k)=\delta t (x_k-0.05)^2$. $\eta_k \sim \mathcal{N}(0,Q_k)$ and $\nu_k\sim \mathcal{N}(0,R_k)$ are process and measurement noise with covariance $Q_k=b^2\delta t$ and $R_k=d^2\delta t$ respectively. The following data are used for simulation: $b=0.5$, $d=0.1$, $\delta t=0.01$ s. The initial truth of state is $x_0=-0.2$, the initial state estimate is $\hat{x}_{0|0}=0.8$ and the initial error covariance is $P_{0|0}=2$. Estimation is performed for time span 0 to 4 s. The system has two equilibrium points, among them it settles one of the stable equilibria either 1 or, -1.  The moderate estimation error force the estimate to settle at wrong equilibrium point, and track loss situation occurs. 
	
\begin{figure}[h!]
		\centering
		\includegraphics[width=8cm,height=5cm]{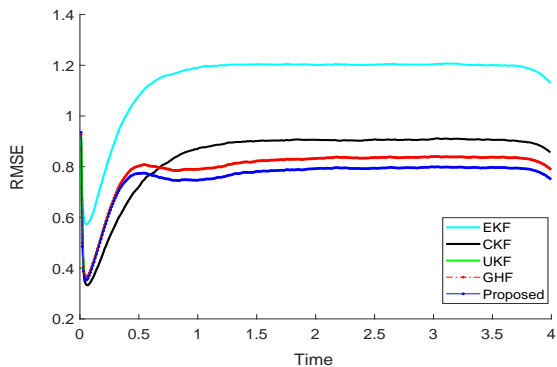}
		\caption{RMSE plot for the EKF, the CKF, the UKF, the GHF and the proposed filter}
		\label{fig_rmse_nl_1d1}
	\end{figure}

     The problem has been solved by the CKF, the UKF, the three points GHF and the proposed filter. The root mean square errors (RMSEs) (excluding fail count, which is discussed later), obtained from 10,000 MC runs  are plotted in Fig. \ref{fig_rmse_nl_1d1}. From the figure, it can be observed that the proposed filter performs better than the EKF, the CKF, the UKF and the GHF.
	
     Filtering performance has also been  compared in terms of fail count. Fail count is the number of run when absolute estimation error is failed to settle below a specified value, say $e_{limit}$, which has been taken 2 here. Fail counts obtained from 10,000 MC runs are summarized in Table \ref{Tab_ito}.  From the Table \ref{Tab_ito}, it can be observed that the proposed filter has the lowest fail count compared to the EKF, the CKF, the UKF, and the GHF. The execution time of all the filters are compared and it has been seen that the run time of the proposed filter is higher (approximately five times)  than other Gaussian filters.
	
	\begin{center}
		\captionof{table}{Percentage fail count and relative execution time taken by different filters}\label{Tab_ito}
		\begin{tabular}{m{2.5cm} m{2.5cm} c }
			\hline
			\hline
			Filter & Fail count (\%) &Execution time \\ 
			\hline
			EKF & 37.59 &0.18 \\
			CKF & 17.55& 0.185\\
			UKF & 14.03&0.19 \\
			GHF & 14.03 & 0.19  \\
			Proposed & 12.53& 1 \\
			\hline
			\hline 
		\end{tabular}
	\end{center}
	
\vspace{0.15cm}
	Problem 2: A BOT \cite{leong2014gaussian} target tracking problem has been considered where a moving target is being tracked from an airborne platform. The engagement scenario has been described in \cite{lin2002comparison}, \cite{sadhu2006sigma} and we omitted here due to page restriction. Readers are requested to see the Figure 1 of \cite{lin2002comparison} to understand the engagement scenario.

	Process model: The discrete time target dynamics on the X-axis is
	\begin{equation*}
	x_{k+1}=\phi_k x_k+\beta_k\eta_k,
	\end{equation*}
	where  
	
	$ x_k = \begin{bmatrix}
	x_{1,k}      \\ 
	x_{2,k}      \\
	\end{bmatrix}$, \hspace{0.5cm} $ \phi_k = \begin{bmatrix}
	1     & T  \\ 
	0      & 1  \\
	\end{bmatrix}$, \hspace{0.5cm}	 $ \beta_k= \begin{bmatrix}
	T^2/2      \\ 
	T      \\
	\end{bmatrix}$,\\
	target position $x_{1,k}$ is in m, target velocity $x_{2,k}$ is in m/s, sampling time $T$ is 0.2 s. Process noise, $\eta_k$, is white Gaussian noise with mean zero and covariance $q=0.01$ $\text{m}^2/\text{s}^4$, and the initial truth of the state is $x_0=[80 \hspace{2mm} 1]^T$. 
	
	Observation model: The platform dynamics in discrete-time is represented as	
	\begin{align*}
	&x_{d,k}=\bar{x}_{d,k}+\Delta x_{d,k} \hspace{0.8cm} k=1,2,\cdots,n_{\text{step}},\\&
	y_{d,k}=\bar{y}_{d,k}+\Delta y_{d,k} \hspace{0.8cm} k=1,2,\cdots,n_{\text{step}} ,
	\end{align*}
	where $\bar{x}_{d,k}$ and $\bar{y}_{d,k}$	are the average platform position in X and Y co-ordinates respectively, $n_{\text{step}}=20$, $\Delta x_{d,k} \sim \mathcal{N}(0,1 \, \text{m}^2)$  and $\Delta y_{d,k} \sim \mathcal{N}(0,1 \, \text{m}^2)$ are assumed to be white, Gaussian and mutually independent noises.  The average platform positions are $\bar{x}_{d,k}=4kT$ and $\bar{y}_{d,k}=20$, where $T=0.2$ s is sampling time. 
	The overall measurement model which includes platform noise \cite{lin2002comparison} can be represented as
	\begin{equation*}\label{BOT_act_meas}
	y_{k}=\tan^{-1}\big(\frac{y_{d,k}}{x_{1,k}-x_{d,k}}\big)+\nu_{k},
	\end{equation*}
	where $\nu_k$ is the measurement noise with mean zero and covariance $R_k$, which is given as
	\begin{equation*}
	E[\nu_k^2]=R_k=\frac{\bar{y}_{d,k}^2+[x_{1,k}-\bar{x}_{d,k}]^2}{\{[x_{1,k}-\bar{x}_{d,k}]^2+\bar{y}_{d,k}^2\}^2}+(\ang{3})^2.
	\end{equation*}
	
\begin{figure}[h!]
	          	\centering
	          	\includegraphics[width=8cm,height=5cm]{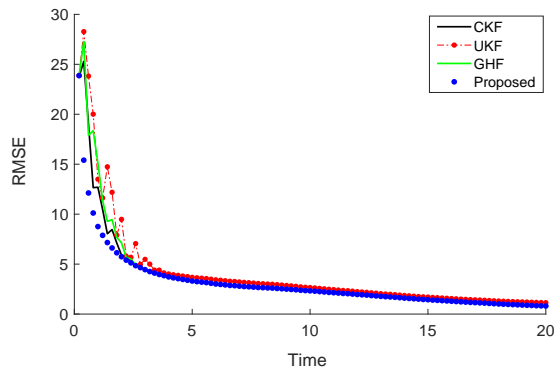}
	          	\caption{RMSE against time plot for position}
	          	\label{fig_rms_bot_pos}
	          \end{figure}

\begin{figure}[h!]
	          	\centering
	          	\includegraphics[width=8cm,height=5cm]{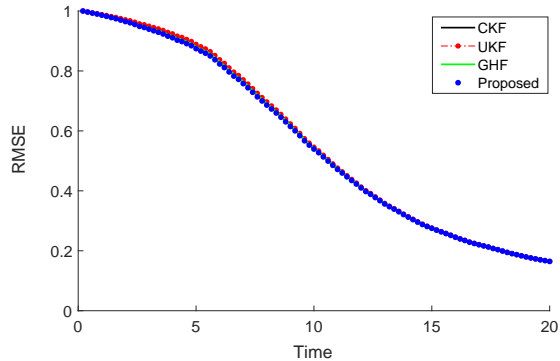}
	          	\caption{RMSE against time plot for velocity}
	          	\label{fig_rms_bot_vel}
	          \end{figure}

	Here, the measurement equation is not in polynomial form, so Taylor series expansion is used. In this problem, third order Taylor series approximation is considered. 
	Initialization of the filter is done as per \cite{lin2002comparison,sadhu2006sigma}. The performance of the filter is shown in terms of RMSE obtained from 10,000 MC runs. The RMSEs of position and velocity (excluding track loss) are shown in Fig. \ref{fig_rms_bot_pos} and Fig. \ref{fig_rms_bot_vel}. From Fig. \ref{fig_rms_bot_pos}, it can be seen that RMSE value for position of the CKF, the GHF and the proposed filter is almost same and slightly better than the UKF. 
	          
	To check the robustness of the estimator against large  initialization error, we vary the initial covariance $P_{0}=\xi P_{0|0}$ (where $\xi$ is a real number $\geq$ 1). The number of track loss, which happens when the last step position estimation error exceeds 15 m, is calculated and tabulated in Table \ref{Tab_BOT}. From the table, we see that the proposed filter has the lowest track loss compared to the CKF, the UKF and the GHF and more robust against initial error.

	The relative execution times of the CKF, the UKF and  the GHF are 0.15, 0.16, and 0.18 respectively with respect to the proposed filter (which is considered to be 1). Here again, we see that the computational cost of the proposed filter is more.	
	\begin{center}
		\captionof{table}{Percentage track loss of different filters}\label{Tab_BOT}
		\begin{tabular}{m{0.5cm} m{1cm} m{1cm} m{1cm} m{1cm} m{1cm} } 
			\hline
			\hline
			& &CKF&UKF &GHF & Proposed  \\ 
			\hline
			$\xi$	&	1 & 0.01  &  0.02 & 0.01 &0\\
			&	5&0.1 & 0.03 & 0.03 &0 \\
			&	7.5 &0.18 &0.08 &0.06 &0.01\\
			&	10 &0.33&0.09 &0.12 &0.01 \\
			\hline		
			\hline
		\end{tabular}
	\end{center}		     			      		
	\section{Discussion and Conclusion}
	Here we propose a new filtering technique, where prior and posterior pdfs are assumed as Gaussian. The intractable integrals encountered in nonlinear filtering are solved by Gaussian integral. If process and measurement functions are polynomial, they directly fit into the proposed filtering framework, if not the Taylor series is used to convert them to a polynomial form. With the help of two examples, it is shown that the proposed filter provides more accurate result than the EKF, the CKF, the UKF, and the GHF in terms of RMSE and fail count. 
	
 \bibliographystyle{ieeetr}
 \bibliography{refer}
\end{document}